\documentclass[twocolumn]{aastex631}
\usepackage{graphicx}
\usepackage{bm}
\usepackage{amssymb}
\usepackage{amsmath}
\usepackage{units}
\usepackage{array}

\usepackage[T1]{fontenc}
\usepackage{aecompl}
\usepackage{calligra}
\usepackage{soul}
\DeclareMathAlphabet{\mathcalligra}{T1}{calligra}{m}{n}
\DeclareFontShape{T1}{calligra}{m}{n}{<->s*[2.2]callig15}{}

\newcommand{\Mdot}[0]{\dot M}

\newcommand{\Mach}[0]{\mathcal{M}}
\newcommand{\Order}[1]{\mathcal{O}({#1})}

\newcommand{\scriptr}{\mathcalligra{r}\,}

\newcommand{\interact}[0]{{\rm int}}

\newcommand{\numag}[0]{\bar \nu}

\usepackage{booktabs}

\usepackage{xcolor}

\submitjournal{ApJ}
\received{October 7, 2024}
\revised{March 12, 2025}
\accepted{March 28, 2025}

\shorttitle{Binaries in truly thin disks}
\shortauthors{Tiede et al.}

\begin{document}

\title{Suppressed accretion onto massive black hole binaries surrounded by thin disks}

\correspondingauthor{Christopher Tiede}
\email{christopher.tiede@nbi.ku.dk}
\author[0000-0002-3820-2404]{Christopher Tiede}
\affiliation{Niels Bohr Institute, Blegdamsvej 17, 2100 Copenhagen, Denmark}
\author[0000-0002-1895-6516]{Jonathan Zrake}
\affiliation{Department of Physics and Astronomy, Clemson University, Clemson, SC 29634, USA}
\author[0000-0002-0106-9013]{Andrew MacFadyen}
\affiliation{Center for Cosmology and Particle Physics, Physics Department, New York University, New York, NY 10003, USA}
\author[0000-0003-3633-5403]{Zolt\'an Haiman}
\affiliation{Department of Astronomy, Columbia University, New York, NY 10027, USA}
\affiliation{Department of Physics, Columbia University, New York, NY 10027, USA}
\affiliation{Institute of Science and Technology Austria (ISTA), Am Campus 1, Klosterneuburg 3400, Austria}
%

\begin{abstract}
    We demonstrate that gas disks around binary systems might deliver gas to the binary components only when the circumbinary disk is relatively warm.
    We present new grid-based hydrodynamics simulations, performed with the binary on the grid and a locally isothermal equation of state, in which the binary is seen to functionally ``stop accreting'' if the orbital Mach number in the disk exceeds a threshold value of about 40. 
    Above this threshold, the disk continues to extract angular momentum from the binary orbit, but it delivers very little mass to the black holes, and instead piles up mass in a ring surrounding the binary. 
    This ring will eventually become viscously relaxed and deliver mass to the binary at the large-scale inflow rate.
    However we show that the timescale for such relaxation can far exceed the implied binary lifetime.
    We demonstrate that the ability of a binary-disk system to equilibrate is dependent on the efficiency at which accretion streams deposit mass onto the binary; which in turn is highly sensitive to the thermodynamic conditions of the inner disk.
    If disks around massive black hole binaries do operate in such non-accreting regimes, it suggests these systems may be dimmer than their single black hole counterparts, but could exhibit dramatic re-brightening after the black holes in-spiral and merge.
    This dimming begins in the UV/optical and could completely choke high-energy emission, such that these systems would likely be intrinsically X-ray weak with reddened continua, potentially resembling the spectra of `Little Red Dots'' recently identified in JWST observations.
\end{abstract}


%
\section{Introduction} \label{S:intro}

Modern cosmology predicts the frequent formation of super-massive black hole binaries (SMBHBs) following major galaxy mergers \citep{Begel:Blan:Rees:1980, Volonteri+2003, Dotti2007, KormendyHo2013}. 
Many of these SMBHBs are expected to interact with gas that reaches the galactic center, much as single super-massive black holes do in ordinary active galactic nuclei (AGN) \citep{Gaskell:1985, BarnesHernquist1996}.
Periodic variability associated with the orbital motion of the binary is a potential signpost of accreting binary massive black holes, and could thus facilitate the selection of SMBHB candidates in time-domain electromagnetic (EM) surveys \citep{Komossa:review:2006, Graham+2015, D'Orazio:PG1302:2015, Charisi:2016, ChenXin:2020}.
SMBHBs might also be multi-messenger sources when they emit gravitational radiation in-band for pulsar timing experiments or future space-based interferometers such as LISA.

Early analytic work on binary accretion posited that tidal torques from the rotating potential would suppress or even halt mass flow onto the binary components \citep{Pringle1991, Milosavljevic2005, LiuShapiro2010}.
Numerical simulations revealed that binaries do in fact expel material from their central region \citep{al94, al96, Cuadra2009, Shi+2012}, but that gas still falls onto the binary via tidal streams \citep{MacFadyen2008, Farris2014, Shi2015}.
It was subsequently demonstrated that the system can come into equilibrium where the accretion rate throughout the disk is constant and equal to that onto the binary \citep{Rafikov2016, MML17, MML19}.

The majority of the numerical work on accreting binaries considers disks with significant pressure support and aspect ratio $h/r = 0.1$ (see \citealt{LaiMunoz:Review:2022} for review).
Canonical disk modeling envisions however, that accretion disks around massive black holes are truly thin with $h/r \sim 10^{-2} - 10^{-3}$ \citep{SS1973}. 
\citet{Ragusa+2016} first treated thinner disks with smoothed particle hydrodynamics (SPH) and reported suppressed accretion rates.
This was supported by studies of finite tori in \citet{Tiede2020} and \citet{Heath+Nixon2020}.
Accretion quenching in colder systems was similarly observed for infinite disks by \citet{Dittmann2022}, but they stressed that the suppression was a spurious artifact of the initial configuration; and that accurately approximating the flow of angular momentum in the disk restores a steady solution (see also \citealt{MML17, Dempsey2020}).

The previous explorations on this topic still considered disks thicker than the canonical thin disk limit for $h/r$ by a factor of a few.
Here we report high-resolution grid-based hydrodynamics simulations approaching the truly thin-disk limit $h/r \sim 10^{-2}$, which --- in conjunction with simple analytic modeling --- show the existence of long-lived non-accreting phases.
We also demonstrate that one can characterize such systems with a single parameter, the stream efficiency (referring to the likelihood that a gas parcel entering the tidally evacuated cavity around the binary becomes bound to one of the black holes).
Our guiding, simplified model for binary accretion is laid out in Section~\ref{S:picture}, and the simulation setups are summarized in Section~\ref{S:setup}.
We report on the dynamics of such truly thin disks in Section~\ref{S:results} and discuss the implications to electromagnetic searches for SMBHBs and future numerical studies in Section~\ref{S:discussion}.
We summarize our main conclusions in Section~\ref{S:summary}.

%
\pagebreak
\vspace{10pt}
\section{Physical picture} \label{S:picture}

A circumbinary disk (CBD) surrounds a binary of mass $M$, semi-major axis $a$, and orbital frequency $\Omega_b$. The large-scale rate of mass flow through the CBD is $\dot M_\infty$, and gas accretes to the binary components at rate $\Mdot_b$. 
The time averaged rate $\langle \Mdot_b \rangle$ and $ \Mdot_\infty $ are equal after a steady-state accretion flow is established.\footnote{$\langle \cdot \rangle$ denotes an average over a viscous time.}
We are considering circumbinary disks that may take a long time to viscously relax, so we do not assume that $\Mdot_\infty$ and $\Mdot_b$ are equal a priori.

The time-varying tidal field of the binary deforms the innermost portions of the CBD, pealing away streams of gas. 
The streams pass near the black holes (BHs), delivering gas onto strongly interacting orbits at a rate  $\Mdot_\interact$. 
A fraction, $\chi \equiv \Mdot_b / \Mdot_\interact$, of the gas parcels lose orbital energy via self-collisions and are gravitationally captured onto orbits around one of the black holes.
They then accrete through one of the circum-single disks called ``minidisks''. 
We refer to this fraction $\chi$ as the stream efficiency.
The remaining gas parcels are tidally up-scattered in the encounter, and rejoin the CBD with their specific angular momentum increased by an amount on the order of $a^2 \Omega_b$.

The gas-induced torque exerted on the binary through this process can be expressed as
\begin{align}
    \dot J_b = (\eta \Mdot_b - \Mdot_\interact) a^2 \Omega_b \, ,
\end{align}
where $\eta = M_1 M_2 / M^2$ is the symmetric mass-ratio. The first term is the accretion torque, and the second term is identified as the tidal torque associated with the rejected, up-scattered gas parcels. The accretion torque is negligible when most of the gas is rejected, $\chi \ll 1$.
In that limit $\dot J_b \simeq - \chi^{-1}\Mdot_b \, a^2 \Omega_b$.
We also define a dimensionless ``torque parameter'', $\ell_0 \equiv \dot J_b / (\Mdot_b a^2\Omega_b)$, representing the torque on the binary per unit accreted mass. 
If a very small fraction of the interacting gas is captured, then $\ell_0 \simeq -1/\chi$, and the quantity $-\ell_0$ may then also be interpreted as the ``recycling number'', meaning the number of times a gas parcel has a strong encounter with the binary before it falls into a black hole.
    \hspace{-5pt}\footnote{$\ell_0 > 0$ has been reported for some system parameters, which arises when $\chi$ is a significant fraction of one. Accurately describing this limit would required an adjustment to our model, but in the $\chi \ll 1$ regime, $\ell_0 < 0$ always.}
When $\chi \ll 1$ the binary interacts with the disk in a ``non-accreting'' phase.
Such a phase would inevitably be temporary, as gas would continue to feed from the large-scale disk and accumulate in a dense "ring" of gas in the inner CBD.
There would be a characteristic timescale $t_{\rm eq}$ for the ring to relax so that a balance is established where $\langle \Mdot_b \rangle \simeq \Mdot_\infty$. 

We consider an accretion flow that cools and flattens  at scales that are large compared to $a$.
To estimate $t_{\rm eq}$, we separate the disk into two zones: an inner zone (the ring) which has adjusted to the binary torque $\dot J_b$, and an outer zone which has not yet become ``aware'' of the inner binary torque \citep[e.g.][]{Kocsis+2012a, Rafikov2016, SBCodeComp:2024, Zrake:CLIs:2025}. 
The disk in the outer zone is formed from the large-scale accretion flow, and carries net-zero radial angular momentum current (a ``torque-free'' disk). 
The ring grows radially over time as the binary influence is communicated outwards \citep{SyerClarke1995, Ivanov99}. The radius at which the two zones meet is called the ``radius of influence'', denoted as $r_\nu(t)$. The radius of influence is defined implicitly as $t_{\rm visc}(r_\nu(t)) = t$, where $t_{\rm visc}(r) \equiv 2r^2 / 3\nu(r)$ is the viscous relaxation timescale at radius $r$, and the viscosity $\nu(r)$ is parameterized explicitly as a function of radius. 

The surface density radial profile in the ring, which is relaxed with respect to the torque parameter $\ell_0$, is
\begin{align}
    \Sigma(r, t) = \frac{\Mdot(t)}{3\pi \nu(r)} \left( 1 - \ell_0 \frac{a^2 \Omega_b}{j(r)} \right) \, ,
 \label{eq:sig-ring}
\end{align}
where $j(r) = \sqrt{GMr}$ is the specific angular momentum of circular orbits at radius $r$ and $\Mdot(t)$ is the mass flow through the ring, and onto the binary. The outer zone (beyond $r_\nu$) maintains the initial configuration $\Sigma_{\rm init}(r) = \Mdot_\infty / 3\pi\nu(r)$. The rate of mass flowing through the inner ring is obtained by imposing a matching condition at the radius of influence $\Sigma(r_\nu) = \Sigma_{\rm init}(r_\nu(t), t)$,
\begin{align}
\label{eq:mdot}
\Mdot(t) &= \frac{\Mdot_\infty}{1 - \ell_0\, a^2 \Omega_b / j_\nu(t)} \\
         &= \frac{\Mdot_\infty}{1 - \ell_0 \left( \frac{3}{2} \numag\, \Omega_b\, t \right)^{-1/3}} \, , \nonumber
\end{align}
where $j_\nu(t) \equiv \sqrt{GM r_\nu(t)}$ is the specific angular momentum at the radius of influence. On the second line we have inserted the $\alpha$-viscosity law $\nu(r) = \numag \sqrt{G M r}$ with $\numag \equiv \alpha\, (h/r)^2$. For a more general viscosity prescription $\nu(r) \propto r^n$, the exponent $1/3$ is changed to $\lambda \equiv (4 - 2n)^{-1}$. For a constant-$\nu$ viscosity law $n=0$ and $\lambda=1/4$.

\begin{figure}[t!]
    \centering
    \includegraphics{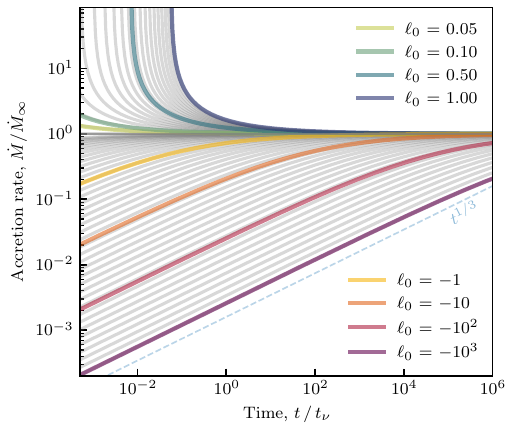}
    \caption{Family of curves for $\Mdot(t)$ from a zero-torque $\alpha$-disk for a given torque parameter $\ell_0$. $t_\nu$ is the viscous time at $r=a$. 
    The dashed-blue line illustrates the power law $t^{1/3}$ growth for an alpha viscosity model in a non-accreting phase.
    }
    \label{fig:mdot-analytic}
\end{figure}

A family of accretion rate curves defined by Equation~\eqref{eq:mdot} is illustrated in Figure~\ref{fig:mdot-analytic} for a range of values of $\ell_0$ in an $\alpha$-disk. When $\ell_0$ is large and negative, $\dot M(t) \propto t^{1/3}$ as shown by the dashed blue line. When $\ell_0$ is positive, the accretion rate is formally positive only at times $t > 2 / (3\, \numag \Omega_b\, \ell_0^{\,3})$. In general this model is not expected to be accurate at very early times. In simulated disks, the early phase is seen to exhibit a complex adjustment phase lasting a fraction of a viscous time, as shown in Figure~\ref{fig:mdot-fits}.

We note that circumbinary disks only permit mildly positive values of $\ell_0 \lesssim \mathcal{O}(1)$. If $\ell_0$ exceeds this range, the density in the ring must approach zero in order to accommodate both the mass and angular momentum flow. This causes the disk to physically detach from the binary system \citep[see][]{Rafikov2016}.
On the other hand, any value $\ell_0 < 0$ is in principle permitted, and thus a non-accreting phase can be long-lived. Equation~\eqref{eq:mdot} can be inverted to express the time required to achieve some accretion rate $\Mdot$,
\begin{align}
    t(\Mdot) = \frac{P_b}{3\pi \numag} \left( \frac{ 1 - \Mdot_\infty / \Mdot}{\ell_0} \right)^{-1/\lambda} \ ,
     \label{eq:nsteady}
\end{align}
where $P_b$ is the binary period. Thus the time required for the disk to equilibrate, i.e. for accretion to ``turn on'', depends strongly on the torque parameter $\ell_0$, or equivalently on the stream efficiency $\chi$. In terms of $\chi$, $t_{\rm eq} \propto \chi^{-3}$ for the $\alpha$-viscosity law and $t_{\rm eq} \propto \chi^{-4}$ for the constant-$\nu$ viscosity law.

%
\section{Numerical Methods} \label{S:setup}
%

%
\subsection{System setup} \label{s:system-and-numerics}

We solved the vertically-averaged Navier-Stokes equations for the evolution of a thin gas disk around a circular, equal-mass binary with the finite-volume code \texttt{Sailfish} \citep{Sailfish:2024}. 
The disk is treated as a 2D viscous fluid in vertical hydrostatic equilibrium with characteristic scale height inversely proportional to the Mach number of the flow $h/r \sim \Mach^{-1}$.
The disk is locally isothermal with sound speed $c_s^2 = -\Phi_g / \Mach^2$ where the potential
\begin{align}
    \Phi_g = \Phi_1 + \Phi_2 \ , \quad \Phi_j = -\frac{GM_j}{\scriptr_j}
\end{align}
with $\scriptr_j \equiv \sqrt{r_j^2 + r_s^2}$ the smoothed distance to binary component $j \in {1, 2}$  for smoothing length $r_s = 0.05\,a$.

The equations of mass and momentum conservation are taken as
\begin{align}
    \partial_t \Sigma &+ \mathbf{\nabla} \cdot (\Sigma\mathbf{v}) = S_\Sigma + b_\Sigma \\
    \partial_t (\Sigma \mathbf{v}) &+ \mathbf{\nabla} \cdot (\Sigma \mathbf{v} \mathbf{v} + P\mathbf{I} - \mathbf{\tau}) = \mathbf{F_g} + \mathbf{S_J} + \mathbf{b_v}
\end{align}
where $v_i$ is the mid-plane fluid velocity, and $P = c_s^2 \Sigma$ is the vertically averaged isothermal gas pressure.
$\mathbf{F_g}$ is the gravitational force density associated with potential $\Phi_g$ and $\tau_{ij} = \nu \Sigma \left( \partial_i v_j + \partial_j v_i - \delta_{ij} \partial_k v_k \right)$ is the viscous stress tensor with $\nu = \alpha c_s h$.
Because our setup does not resolve the binary components down to their accretion boundaries, the source terms $S_{\{\Sigma, J\}}$ are mass and momentum sinks (respectively) meant to mimic accretion onto each binary component.
We employ spinless sinks with characteristic radius $r_s$ and removal timescale $\Omega_b^{-1}$ (see \citealt{Dempsey2020, Dittmann+Ryan2021} for details).

The outer boundary is set to enforce a steady accretion rate $\Mdot_\infty$ via a buffer term $b_{\{\Sigma, \mathbf{v} \}}$ that drives the solution back towards the initial condition (see e.g. \citealt{Westernacher-Schneider:2022}).
This allows the solution to mimic that of an infinite accretion disk with a constant viscous inflow of material and neutralizes artifacts associated with the square domain.

\begin{figure*}[h!]
    \centering
    \includegraphics{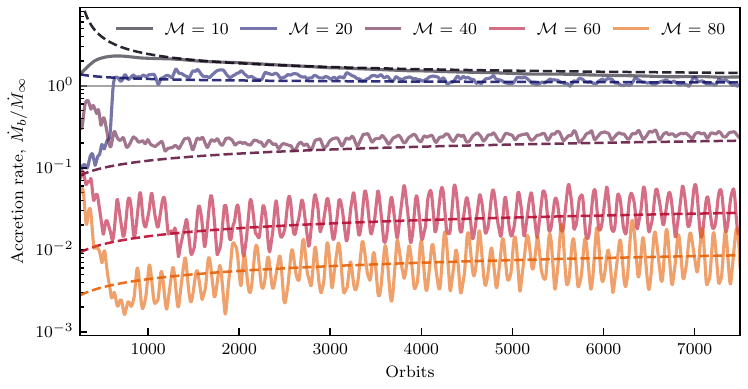}
    \caption{
     Accretion rates for each system from the ``standard'' setup. Dashed lines are from the analytic expression for $\Mdot(t)$ from Equation~\eqref{eq:mdot}.
     The torque parameter is determined empirically as $\ell_0 = \{0.59, 0.17, -7.05, -66.13, -224.22\}$ (in order of growing $\Mach$) from averages over a viscous time ($\sim 4\times 10^3\,\unit{orbits}$ at the cavity edge). 
     We note that the $\Mach = 80$ run is not a converged result (as will be shown in Figure~\ref{fig:stream-efficiency}), but that the accretion rate is still fully consistent with the measurement of $\ell_0$ at this resolution.
    }
    \label{fig:mdot-fits}
\end{figure*}

%
\subsection{Simulations} \label{s:setup-and-simulations}

The fiducial disks are circular and Keplerian with initial surface density profiles corresponding to a torque-free disk with an excised cavity, $\Sigma_{\rm init} \times f_{\rm cav}$ . 
The cavity function $f_{\rm cav}(r) = \delta + (1 - \delta) \exp{\left[-(2.5\, a / r)^{12} \right]}$, $\delta = 10^{-5}$ allays the start-up transient and expedites the approach to steady-state \citep{MacFadyen2008}.
This is the ``standard setup'' from \citet{SBCodeComp:2024} modified for an $\alpha$-viscosity.
Note that in this standard setup (with $\Mach = 10$ and $\nu = 10^{-3} a^2\Omega_b$), the gas-induced binary torque is positive, and the accretion rate over-shoots the inflow rate $\dot M_\infty$ while the disk relaxes from a zero-torque configuration. 
In scenarios where the gas-induced binary torque is negative, we expect the opposite effect, as discussed in Section~\ref{S:picture}.
Numerical solutions in these scenarios will be valid until the radius of influence becomes of order the box size ($\sim 10^5$ orbits) and the relaxed inner solution begins to conflict with the outer-boundary condition.

Because the coupling to the binary induces a radial angular momentum current, in practice a true steady-state can only be realized by matching this current at the outer boundary.
This can be realized by initializing the surface density profile according to Equation~\eqref{eq:sig-ring} and guessing the torque parameter $\ell_0$ (see e.g. \citealt{MML17, Dempsey2020, Dittmann2022}).
We follow this procedure in Section~\ref{s:steady-states} to determine the steady-state behavior of the considered systems.

We explore solutions for increasingly cold disks with $\Mach = \{10, 20, 40, 60, 80, 100\}$. 
To control against varying viscous torques, we fix $\numag = 10^{-4}$.
This yields values $\alpha = \{0.01, 0.04, 0.16, 0.36, 0.64, 1.0\}$, respectively, and fixes the local viscous times $t_\nu \sim (\numag \Omega_b)^{-1} \approx 10^3$ orbits.
The simulation domain is a square of total side length $24\,a$. To probe numerical convergence, the grid resolution is varied for each run in the range $\Delta x = 0.004\,a -  0.01\,a$.
%
%

%
\section{Results} \label{S:results}
\subsection{The non-accreting phase} \label{s:non-accrete-phase}

Figure~\ref{fig:mdot-fits} shows the accretion rates for the standard systems evolved at resolution $\Delta x = 0.01\,a$ for approximately two viscous times at the cavity edge $\gtrsim 8000$ orbits.
We observe that this configuration yields a quasi-steady configuration for the $\Mach \leq 20$ systems, but that all other solutions are still in a non-accreting phase with $\Mdot_b \ll \Mdot_\infty$.
We additionally include the analytic estimate from Equation~\eqref{eq:mdot} (where we have taken $\Mdot = \Mdot_b$) as dashed lines of the same color, with $\ell_0$ determined empirically as $\langle \dot J_b \rangle / (\langle \Mdot_b \rangle \, a^2 \Omega_b)$.
We find that Equation~\eqref{eq:mdot} captures the normalization and evolution of $\Mdot_b$ remarkably well for both the over- and under-accreting systems.
Use of Equation~\eqref{eq:nsteady} and the measured values of $\ell_0$ imply that the $\Mach = 40$ system would require $\Order{ 10^5}$ orbits to reach $\Mdot_b = 0.5 \Mdot_\infty$, while the $\Mach = 80$ system would require $\Order{10^{10}}$ orbits.

%
\subsection{Steady-states} \label{s:steady-states}
%

\begin{figure*}[t!]
    \centering
    \includegraphics{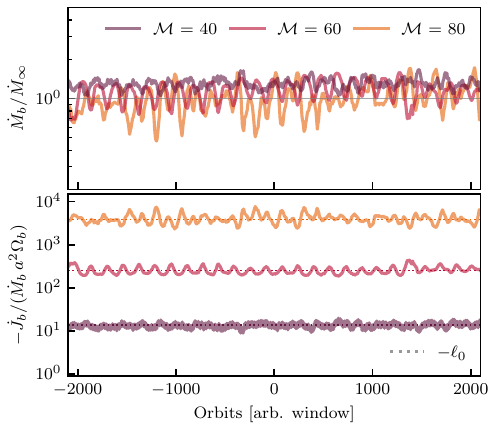}
    \includegraphics{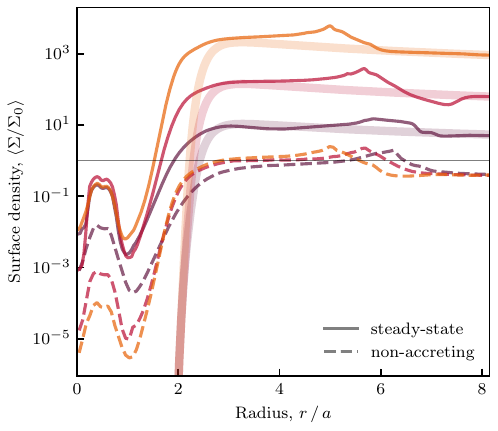}
    \caption{(Left) Time-series of an arbitrary window of the binary accretion rate and torque per accreted mass in an approximate steady-state. 
    The dotted lines denote the measured torque parameter, $-\ell_0$.
    (Right) The average surface density profiles from the steady-state phase in solid, and the non-accreting phase in dashed.
    $\Sigma_0$ is the characteristic disk density $\Mdot_\infty / 3 \pi \numag$, and the initial profiles for the steady-state runs are illustrated by the faint bands of equivalent color.
    }
    \label{fig:steady-state}
\end{figure*}

Given sufficient time and mass, any disk-binary system will ultimately achieve a steady-state.
However Equation~\eqref{eq:nsteady} shows that it could be prohibitive to run a simulation to $t_{\rm eq}$.
Nevertheless, a simulation can be prepared near a steady state by initializing the CBD with the surface density profile in Equation \ref{eq:sig-ring} and guessing the torque parameter $\ell_0$ \citep[][]{MML17, Dempsey2020, Dittmann2022}.

It turns out the torque parameter can be accurately measured even when the system is not in a steady state.
This is a main result and is detailed in the following section (Section~\ref{S:self-similar-evo}).
At a higher resolution $\Delta x = 0.008\,a$, we measured $\ell_0$ from the standard setup over a viscous time (these can be seen in Figure~\ref{fig:stream-efficiency}).
We then started new simulations for $\Mach \geq 40$ with these values of $\ell_0$.
We denote this set of initial conditions as $\Lambda$.
Starting from $\Lambda$, after the start-up transient we observed the binary to accrete in a statistical steady-state with $\langle \dot M_b(t) \rangle \simeq \dot M_\infty$ and $ \dot J_b / (\Mdot_b\, a^2 \Omega_b) \simeq {\rm const}$, as seen in the left panels of Figure~\ref{fig:steady-state}.
The dotted line illustrates the associated value $-\ell_0$.

The right panel of Figure~\ref{fig:steady-state} shows the average surface density profiles from both the non-accreting phase (dashed lines; Section~\ref{s:non-accrete-phase}) and the steady-state phase (solid lines) in units of the characteristic disk density $\Sigma_0 = \Mdot_\infty / (3 \pi \numag a^2 \Omega_b)$.
The faint bands illustrate the initial profiles for the steady-state runs.
The steady-state surface density of the inner disk is enhanced by an approximate factor of $-\ell_0$, and the average radial velocity in the inner disk is suppressed by a similar factor in order to accommodate the large radial angular momentum current.
The equilibrium minidisk density also appears insensitive to $\Mach$ (similarly observed in \citealt{Tiede2020}).
This is possibly because it adjusts itself to accommodate $\Mdot_\infty$, and thus, is sensitive to viscosity $\nu(\alpha)$, but not $\Mach$ (when fixing the strength of viscous stresses).
Conversely, in the non-accreting phases the minidisks are starved with surface densities proportional to $\Mdot_b$.

%
\subsection{Self-similar evolution} \label{S:self-similar-evo}
%

\begin{figure}[t!]
    \centering
    \includegraphics{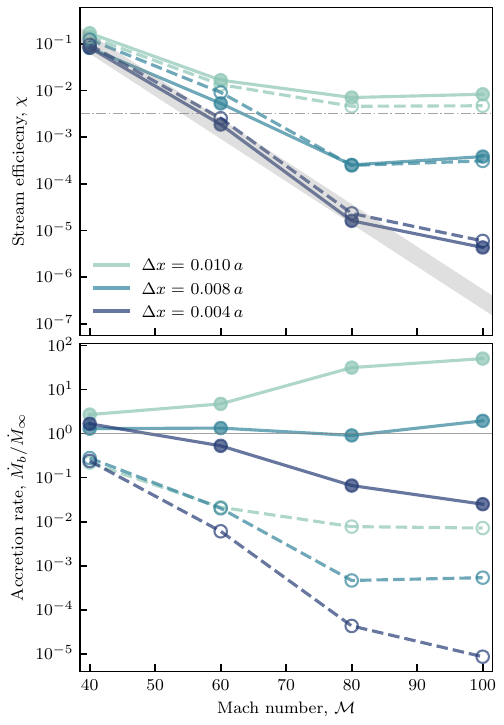}
    \caption{
    Stream efficiency $\chi \approx \ell_0^{-1}$ (top) and accretion rate (bottom) as a function of Mach number and resolution. 
    Dashed lines denote the ``standard'' initial disk configuration \protect\citep{SBCodeComp:2024}.
    The solid lines indicate initial condition set $\Lambda$, which permitted a steady-state at $\Delta x = 0.008\,a$.
    $\Mdot_b$ is sensitive to the surface density profile of the system but $\chi$ is not.
    The dashed-dotted line denotes the condition from Equation~\eqref{eq:turn-on} below which a binary will merge before accreting significantly. 
    The light-grey band illustrates the exponential function $\chi(\Mach)$ in Section~\ref{s:obs-implications}.
    }
 \label{fig:stream-efficiency}
\end{figure}

\begin{figure*}[t!]
    \centering
    \includegraphics{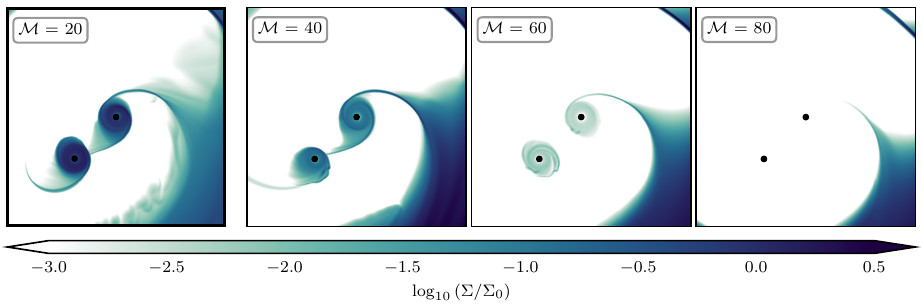}
    \vspace{-8pt}
    \caption{Density snapshots during stream formation from the ``standard setup'' of Seciton~\ref{s:non-accrete-phase}. The left-most panel for $\Mach = 20$ is in steady-state, while the other three are under-accreting $(\Mach = 40)$ or in a non-accreting phase ($\Mach \geq 60$).
    }
    \label{fig:stream-widths}
\end{figure*}

We stress that the results of Sections~\ref{s:non-accrete-phase}~and~\ref{s:steady-states} represent two phases of evolution for the same systems.
The former a transient (albeit potentially long-lived) phase as the disk adjusts to angular momentum injection from the binary, and the latter the equilibrium configuration which persists either until the outer gas reservoir is depleted or the binary merges.

It is not self-evident that the stream efficiency, and in turn $\ell_0$, is insensitive to the varying gas distribution (e.g. Figure~\ref{fig:steady-state}, right).
However, we find that when $\chi$ is small, this is the case and that $\chi \simeq -\ell_0^{\, -1}$ is generally a fixed property of the system
(i.e. does not change over time as the system equilibrates).
Figure~\ref{fig:stream-efficiency} shows the stream efficiency (top panel) determined as $\chi = \ell_0^{\,-1}$ from the results of both Sections~\ref{s:non-accrete-phase}~and~\ref{s:steady-states}.
Because these had different resolutions, we supplement each with 2000 orbits of their complimentary simulations.
Namely, \emph{dashed lines} denote systems initialized in the standard setup, and \emph{solid lines} those with initial condition set $\Lambda$ (from Section~\ref{s:steady-states}).
We additionally include ``upsampled'' solutions that have had their resolution doubled {\bf to $\Delta x = 0.004\,a$} and evolved for an additional few hundred orbits.
To emphasize how the solid- and dashed-lines symbolize dramatically dissimilar accretion states, the bottom panel of Figure~\ref{fig:stream-efficiency} shows the associated accretion rates.
The stream efficiency and torque parameter are insensitive (at leading order) to the mass distribution and to whether the binary is overfed, starved, or in steady-state.

The critical feature for determining $\chi$ is the system resolution.
The growing separation between solid-dashed pairs (each color) in the top panel also indicates a lack of numerical convergence. 
We posit that this decrease in stream efficiency---and its resolution dependence---occurs because the streams become increasingly thin and sharp as the disk temperature is reduced.
For ballistic particles pulled into streams, the tidal effect is to induce orbit crossing \citep{Dorazio2016, Tiede2022}.
Thus, gravity tries to compress the stream into an infinitely thin filament, and the width of the stream is only supported by the transverse pressure gradient. 
As the pressure support is decreased with increasing Mach number, the streams narrow and begin to miss the binary minidisks, decreasing their per-stream mass deposition on the binary.
In support of this hypothesis, we plot in Figure~\ref{fig:stream-widths} surface density snapshots of stream formation at equal orbital phase for four $\Mach$ considered.
One can appreciate the diminishing apparent thickness with growing Mach number.
We posit that the sensitivity to resolution occurs when the natural stream width falls below the resolution scale. 
This artificially truncates the compression of the stream yielding a spuriously large value for $\chi$.
This effect would additionally account for the apparent plateaus in $\chi(\Mach)$ for $\Mach \gtrsim 80$ at each resolution.
Until the simulation is numerically converged, $\chi$ decreases monotonically with increasing resolution.

Generally, $\chi$ decays extremely rapidly with disk temperature, and as one approaches the continuum limit the stream efficiency can become as low as $\chi < 10^{-5}$ as $\Mach \rightarrow 100$.
We thus expect that $\chi$ would be lower than $10^{-5}$ in a numerically converged run of a truly thin disk with $\Mach \sim 100 - 1000$, and we conclude that circumbinary black hole accretion disks in nature might persist in non-accreting configurations over timescales exceeding the binary lifetime.

%
\section{Discussion} \label{S:discussion}
%

%
\subsection{Numerical modeling and physical caveats}
\label{s:modeling-future}

As noted in the Introduction, comparatively mild suppression of the binary accretion rate in increasingly cold disks had previously been observed in both finite volume and SPH studies.
However, it had either been attributed to an associated decrease in the strength of viscous torques or regarded as a spurious manifestation of a poorly tuned initial condition.
We have demonstrated not only that non-accreting phases can be long-lived relative to the binary lifetime, but that they occur at fixed viscous strengths.~\hspace{-5pt}\footnote{For the careful reader we also note that there is an additional viscosity dependence in the determination of $\chi(\Mach;\, \numag) \approx \ell_0^{-1}$, where a lower viscosity would decrease $\chi$ and vice versa \citep[c.f.][]{Dittmann2022}.}

Such non-accreting phases occur because the efficiency of mass capture from streams drops rapidly with decreasing disk temperature (Section~\ref{S:picture}).
However, we have determined that $\chi$ is insensitive to the system state, and only depends on the hydrodynamic properties of the disk and resolution.
This is consistent with the observation in the literature that $\dot J_b(t) / \Mdot_b(t)$ seemingly always equilibrates to a stable value (close to the steady-state solution) in a viscous time or less;
even in the presence of a time-evolving gas distribution and accretion rate 
(e.g. because of a finite disk that is being depleted \citealt{Munoz19, Tiede2020, Penzlin2022}).

From Figure~\ref{fig:stream-efficiency} and Section~\ref{S:picture} it is clear that the accretion rate, equilibration time, steady-state angular momentum current, and equilibrium gas-morphology are highly sensitive to the temperature of the circumbinary material, and particularly of the material that strongly interacts with the binary in streams.
Thus, each of these quantities could in turn vary sensitively with physics that alter the inner disk vertical structure and thermodynamics.
This may include a full treatment of heat injection and cooling mechanisms (e.g. \citealt{Westernacher-Schneider:2022, WangBaiLai:2023, Cocchiararo:2024}), binary eccentricity \citep{Zrake2020, Dorazio2021, Siwek2023}, the inclusion of magnetic fields (c.f. \citealt{Shi2015, Hopkins:MAD-Quasars:2024, Most:MHD:2024}), or radiative effects \citep{Luciano2018, Williamson2022} in such truly thin conditions.
Moreover, while treating the problem in two dimensions may well be valid for such thin solutions, warm circumbinary disks studied in 3D typically posses less eccentric and smaller cavities \citep[e.g.][]{Shi+2012, Shi2015, Moody19, SBCodeComp:2024, Most:MHD:2024}.
These are more closely attached to the binary and may permit larger stream efficiencies, even in such cold, isothermal solutions.
Quantifying how additional physics may alter the inner mass flow, stream capture dynamics, and in turn the equilibrium angular momentum current, then, will be crucial for more reliably determining the feasibility of discovering compact massive binaries as multi-messenger sources and for accurately determining their associated characteristics.

%
\subsection{Observational Implications}
\label{s:obs-implications}

While there are many physical processes that could meaningfully alter inner-disk dynamics, we examine the possible observational ramifications of such long-lived non-accretion phases.

\paragraph{Binary lifetimes}
We have demonstrated that massive binaries embedded in realistically thin disks may undergo sustained phases of negligible accretion.
Despite the truncation of the mass flow in the non-accreting phase, gas parcels continue to interact with the binary removing angular momentum with each strong encounter.
To accommodate the sustained injection of angular momentum, the mass of the inner ring grows with time.
The characteristic timescale for the binary to merge then is $t_m \sim M / \Mdot_\infty$, that required for the disk to supply an equivalent binary mass to the inner-ring.
For a disk with $\Mdot_\infty$ taken as the Eddington rate, this is a Salpeter time $t_{\rm sal} \simeq 4.5 \times 10^7\,\unit{years}$.
To compare with $t_{\rm eq}$, we consider $\chi(\Mach)$ as a simple decaying function $ k_0 e^{- k_1 \Mach}$.
This relation is shown in Figure~\ref{fig:stream-efficiency} as a light grey band for $k_0 = 440$ and $k_1 = 0.21$.
For thin, radiatively efficient (isothermal) accretion, this is likely a conservative estimate as we anticipate the dependence to steepen as resolution is increased.
One can readily see that the decaying $\chi(\Mach)$ causes $t_{\rm eq}$ to rapidly overtake $t_{\rm sal}$ with growing Mach number.
 Because the ``turn on'' time is dominated by this relation, it can be approximated as 
\begin{align}
    t_{\rm eq}& \simeq 1.3 \times 10^{14} \, \left( \frac{P_b}{1\,\unit{yr}} \right) \times  \exp{\left[ 0.6 \left( \frac{\Mach_{3 a}}{80} \right) \right]}\; \unit{yr}
\end{align}
where
\begin{align}
\nonumber
    \Mach_{3 a} \simeq 80 \left( \frac{\alpha}{0.005} \right)^{\! 1/10} \!\left( \frac{f_{\rm Edd}}{1.0} \right)^{\! -1/5} \!\left( \frac{P_b}{1\,\unit{yr}} \right)^{\! -1/30} \!\left( \frac{M}{10^5\, {\rm M_\odot}} \right)^{\! 2/15} 
\end{align}
is the Mach number in the inner ring ($r = 3\,a$) in a gas-pressure dominated disk with primary opacity from electron scattering accreting at $\Mdot_\infty$ equal to $f_{\rm Edd}$ times the Eddington rate. 

One can also compute a general stream efficiency below which a binary will merge before it ``turns on'' (i.e. $\Mdot_b = 0.1 \Mdot_\infty$) by equating $t_{\rm eq}$ and $t_{\rm sal}$.
This yields a critical value
,
\begin{align}
 \label{eq:turn-on}
    \chi_c \simeq  0.003 \; \bigg( \frac{\numag}{10^{-4}} \bigg)^{-1/3} \bigg( \frac{P_b}{1\,\unit{yr}} \bigg)^{1/3}  .
\end{align}
This condition is illustrated as a dashed-dotted line in the top-panel of Figure~\ref{fig:stream-efficiency}.
For shorter initial periods, the gravitational wave inspiral time also generally becomes shorter than $t_{\rm eq}$.

\paragraph{Time-domain searches}
Such suppressed accretion phases and severe ``turn on'' times could pose challenges for searches for compact massive binaries in electromagnetic surveys.
One of the primary search techniques is to identify periodic emission features tied to the underlying binary period.
The hydrodynamic component of such periodicity is typically connected with the binary accretion rate \citep[e.g.][]{Farris2014, D'Orazio:PG1302:2015, Combi2021, Westernacher-Schneider:2022}.
The most prominent variability is also expected to be associated with higher energy emission (like in X-rays or the UV) from the binary minidisks \citep{Cocchiararo:2024}.
These components may not in fact be present under sustained suppression of the binary accretion rate and the associated starvation of the minidisks.
Alternative search methods based on the periodic lensing or boosting of emission from material co-moving with the binary components \citep{D'Orazio:nature:2015, D'Orazio:Lensing:2018, DavelaarHaiman:2022} are also likely to be compromised in the absence of persistent minidisks.

Variability at lower frequencies (e.g. in the optical or infrared) is typically associated with features in the inner disk \citep{Westernacher-Schneider:2022} and may yet be present during non-accreting phases (infrared variability may also come from reverberation \citealt{D'Orazio:Lighthouse:2017}).
However, the truncation of the flow against the binary tidal barrier at $\sim {\rm few} \times a \gg GM / c^2$, as opposed to near the inner most stable circular orbits may considerably lower the total luminosity of such sources.
The luminosity suppression will be of order $a / r_g$ where $r_g = GM / c^2$ is the gravitational radius.
For massive binaries with month-to-year long orbital periods, this is $a / r_g \sim \Order{10^2 - 10^4}$ which could make binaries with truly thin disks meaningfully dimmer and harder to discover than their single-BH AGN-counterparts.

That said, regardless of whether the system ``turns-on'' or not, a reservoir of mass approximately equal to that of the binary will be released following a GW-driven merger.
This could instantiate periods of high-Eddington accretion in a ``dam-break'' scenario and cause the post-merger black hole to light up as a luminous quasar \citep{Milosavljevic2005, Shapiro:2010, Tanaka:2010}.

\paragraph{Spectral characteristics.}
The truncation of the circumbinary accretion flow and suppression of the minidisks will also alter the spectral appearance of accreting binaries.
Massive binaries in such non-accreting phases may posses some classical features of AGN, but may also present as non-standard AGN types.
In particular, if one assumes that thermal emission from the disk at frequency $f$ is primarily emitted at the radius where $k T(r) \sim h f$, the truncation of the accretion flow will censor emission above some characteristic frequency $f_a$ where $r(f_a) \approx a$ \citep{Roedig+2014}.
The characteristic temperature associated with $r(f_a)$ is
\begin{align}
    T_a = \left( \frac{3}{8 \pi \sigma} \frac{G M \Mdot}{a^3} \right)^{1/4}
\end{align}
giving
\begin{align}
    k T_a \simeq 2.5 \, \left( \frac{f_{\rm Edd}}{1.0} \cdot \frac{M}{10^7 {\rm M_\odot}} \right)^{1/4} \left( \frac{P_b}{1\,\unit{yr}} \right)^{-1/2}\;\unit{eV}
\end{align}
This reference system would be censored in the blue optical and appear reddened compared to an untruncated disk (potentially mimicking the effects of dust attenuation).
At shorter periods (or larger masses / accretion rates), this cutoff would move into the UV. 
For longer periods (or lower masses / Eddington fractions) the disk truncation could suppress the optical emission entirely, and possibly prevent the formation of broad-line emission regions characteristic of standard AGN.

Sustained starvation of the minidisks will also eliminate the hottest regions near the BH horizons that source X-ray emission. 
At mildly suppressed accretion rates  each black hole may retain a hot, radiatively inefficient component that would radiate X-rays through Bremsstrahlung or inverse Compton scattering of photons from the circumbinary continuum.
At the extreme accretion suppressions implied for $\Mach \gtrsim 60$ from the highest resolution runs in Figure~\ref{fig:stream-efficiency}, however, the circum-single flow is likely so rarefied that it would produce very little high energy emission.
Such configurations would be intrinsically X-ray weak.

The primary spectral signpost of a binary in a non-accreting phase would then be intrinsic X-ray weakness in combination with a reddened (and possibly suppressed) optical/UV continuum.
An interesting class of objects which roughly fit this qualitative description are the recently discovered population of high-redshift galaxies dubbed the ``Little Red Dots'' (LRDs;~\citealt{Matthee+2024-LRD}). 
LRDs have reddened rest-frame optical spectra \citep{Labbe:LRDs:Nature:2023} and are generally undetected in X-rays
\citep{Yue:LRDs:2024}.~\hspace{-5pt}\footnote{Another fundamental feature of LRDs is a blue excess in the rest-frame UV, but this may be sourced by galactic star formation instead of the AGN disk \citep{Kocevski:LRDs:2023, Greene:LRDs:2024}.}
Whether or not a binary in a non-accreting phase can self-consistently account for the specific nuances of LRD spectral energy distributions, though, requires more detailed modeling and follow up.
%
%

%
\section{Summary} \label{S:summary}

We have studied the dynamics of massive binaries accreting from disks approaching the truly thin limit, $\Mach \rightarrow 100$.
We have determined that such systems are highly inefficient at delivering mass to the binary, 
and as a result may undergo exceptionally long-lived non-accreting phases.
We find that the time evolution of such systems can be well described with a simple one-parameter model where the critical quantity is the stream efficiency $\chi$.
This sets the characteristic equilibration time for the system, the equilibrium angular momentum current in the disk, and the associated steady-state density distribution. 

Moreover, we have shown that $\chi$ is an intrinsic property of the system which does not vary (to leading order) with the system accretion state or gas distribution.
It is, however, highly sensitive to numerical spatial resolution. 
For gas conditions where we were unable to obtain numerically converged measures of $\chi$, it trends toward smaller values with increasing grid resolution, indicating that our simulations have only set upper bounds on $\chi$ as a function of Mach number.
As a result, we posit that if disks around massive binaries are well described by such a truly thin, isothermal configuration, then they may undergo substantial phases of limited or no accretion.
This may severely limit the prospects for discovering such systems in electromagnetic surveys.
It is important to mention that our estimates for the duration of non-accreting phases might only be accurate when the accretion flow forms an extended disk around the binary; if instead low-angular momentum gas were to fall more directly toward the binary, it might cause the formation of a dense ring of gas that could be accreted more promptly.

Lastly, we have determined that the stream efficiency and all subsequent properties of the system are exceptionally sensitive to the thermodynamic conditions of the inner disk.
It is crucial both for modeling of SMBHB populations, and for the identification of SMBHB candidates from time-domain EM surveys, to understand whether realistically cold circumbinary disks exhibit such low stream efficiencies when 
additional physics are included.

%
\begin{acknowledgements}
C.T. sincerely thanks Daniel J. D'Orazio for useful and illuminating discussions.
This work was supported by the European Union’s Horizon 2023 research and innovation program under Marie Sklodowska-Curie grant agreement No. 101148364, 
by Sapere Aude Starting grant No. 121587 through the Danish Independent Research Fund,
by the LISA Preparatory Science Program (LPS) through NASA grant 80NSSC24K0440, 
and by NASA Astrophysics Theory Program (ATP) grant 80NSSC22K0822. 
Computation time for this work was supported through the NYU IT High Performance Computing resources as well as the Tycho supercomputer hosted at the SCIENCE HPC center at the University of Copenhagen.
\end{acknowledgements}

%
\bibliographystyle{mnras}
\bibliography{refs}

\end{document}